\documentclass[%
     reprint,
     amsmath,amssymb,
     aps,
]{revtex4-1}

\usepackage{graphicx}
\usepackage{dcolumn}
\usepackage{bm}

\begin{document}
\title{Evolutionary dimension reduction in phenotypic space}
\author{Takuya U. Sato}
\author{Kunihiko Kaneko}%
\email{kaneko@comple.c.u-tokyo.ac.jp}
\affiliation{Graduate School of Arts and Sciences, University of Tokyo\\ 3-8-1, Komaba, Meguro-ku, Tokyo 153-0041, Japan}
\date{\today}
\begin{abstract}
In general, cellular phenotypes, as measured by concentrations of cellular components, involve large degrees of freedom. 
However, recent measurement has demonstrated that phenotypic changes resulting from adaptation and evolution in response to environmental changes are effectively restricted to a low-dimensional subspace. 
Thus, uncovering the origin and nature of such a drastic dimension reduction is crucial to understanding the general characteristics of biological adaptation and evolution.
Herein, we first formulated the dimension reduction in terms of dynamical systems theory: considering the steady growth state of cells, the reduction is represented by the separation of a few large singular values of the inverse Jacobian matrix around a fixed point.
We then examined this dimension reduction by numerical evolution of cells consisting of thousands of chemicals whose concentrations determine phenotype. 
The model cells grow with catalytic reactions governed by genetically determined networks, which evolve to increase cellular fitness, i.e., growth speed.
As a result of the evolution, phenotypic changes due to mutations and external perturbations were found to be mainly restricted to a one-dimensional subspace.
One singular value of the inverse Jacobian matrix at a fixed point of concentrations was significantly larger than the others.
The major phenotypic changes due to mutations and external perturbations occur along the corresponding left-singular vector, which leads to phenotypic constraint, and fitness dominantly changes in the same direction.
Once such phenotypic constraint is acquired, phenotypic evolution to a novel environment takes advantage of this restricted phenotypic direction.
This results in the convergence of phenotypic pathways across genetically different strains, as is experimentally observed, while accelerating further evolution.
We also confirmed that this one-dimensional constraint on phenotypic changes is imposed even by evolution under fluctuating conditions with environmental changes occurring every few generations, where the fitness for each condition is embedded into the evolving one-dimensional direction for major phenotypic changes.
Thus, while genetic evolution can be random, phenotypic evolution appears to be constrained.
\end{abstract}
\maketitle

\section{INTRODUCTION}
    Biological systems generally contain large degrees of freedom. 
    Cells contain a huge variety of components and grow as a result of their complex reaction dynamics.
    Even proteins, which are relevant to biological function, consist of a large number of units.
    Currently, several methods have been developed to extract such high-dimensional data from cells, including transcriptome, proteome, and metabolome analyses to measure the abundances of chemicals within a cell \cite{Matsumoto2013GrowthBacteria,Schmidt2016TheProteome,Marguerat2012QuantitativeCells}.
    However, in spite of advances in these omics measurements, extracting biologically important information from such high-dimensional data remains difficult.
    
    Biologically important information, such as cell growth rate, capacity to adapt to novel environmental conditions, and survivability under stressful conditions, is often expressed in terms of a few degrees of freedom of variables.
    If a few relevant variables could be extracted from many intracellular variables, it may be possible to bridge high-dimensional omics data with biologically essential information.
    Of course, it could be too optimistic to assume that such a reduction to a few variables is possible for all cellular states, as these include dynamically changing states during differentiation and dormant states upon nutrient depletion.
    Instead, by restricting our interest to cells growing while preserving their intracellular compositions under sufficient nutrient supplies, we may be able to achieve the desired dimension reduction.
    
    Indeed, in cellular states with such steady exponential growth, several laws represented by a few variables or parameters have been uncovered.
    These include the classic laws by Monod \cite{Monod1949} and Pirt \cite{Pirt1965,Pirt1982}, whereas Scott and Hwa recently unveiled proportionality between changes in growth rate and ribosome abundance \cite{Scott2010,Scott2011,Scott2014} following the observations of Schaechter et al. \cite{Schaechter1958}.
    Furthermore, by noting that all components are equally diluted by steady cell growth, common proportionality in the changes of chemical concentrations across all components can be assumed, as has been experimentally verified by transcriptome and proteome analysis over thousands of mRNA and protein species.
    Such analyses have shown that the common proportion coefficient in gene expression changes agrees with that in cell growth rate \cite{Kaneko2015,Furusawa2015b,Furusawa2018}.

    Similar relationships have also been recently uncovered in laboratory evolution of bacteria \cite{Furusawa2015b}.
    Changes in gene expression induced by evolution and adaptation to environmental changes are highly correlated across thousands of genes. 
    
    The common proportionality of expression changes upon adaptation and evolution suggests that even though cells involve a great number of components, adaptive changes in cellular states are restricted to a lower-dimensional subspace.
    Cell model simulations with stochastic catalytic reaction dynamics have shown that that adaptive changes in chemical concentrations after evolution are restricted to the one-dimensional sub-space of chemical composition \cite{Furusawa2015b, Furusawa2018}.
    Furthermore, repeated bacterial evolution experiments have suggested that changes in chemical concentrations follow the same low-dimensional paths despite differences in genetic changes\cite{Horinouchi2017PredictionMutations, Horinouchi2015b}.
    The hypothesis that evolution to increase fitness leads to this dimension reduction seen in phenotypic change explains the experimental results of common proportionality.
    
    However, how this dimension reduction to a few degrees of freedom occurs through evolution is unclear.
    Can the reduction be formulated explicitly in terms of dynamical systems?
    Note that the dimension reduction suggests that phenotypic variation in one- or few-dimensional subspace is much larger than those in other dimensions, or in other words, the relaxation process along the subspace is much slower than in the other dimensions.
    If this is the case, can one characterize the dimension reduction of phenotypes as a separation of slow modes?
    Last, but not least, can the convergence observed in phenotypic evolution mentioned above be explained in terms of the dimension reduction of phenotypic changes?

    In the present paper, we address these questions by formulating the dimension reduction as an emerging property of the relaxation spectrum to a steady state as a result of evolution.
    By computing the Jacobian matrix for the relaxation dynamics to the steady state, we demonstrate that one singular value is separated from others and trends closer to zero through evolution. 
    We then discuss how environmental switches may influence this dimension reduction.

    In Sec.\ref{sec:theory_of_dimension_reduction}, we describe the formulation of the dimension reduction of phenotypic changes in terms of dynamical systems theory.
    The dimension reduction is formulated as the separation of larger singular values of the inverse Jacobian matrix in relaxation dynamics of the cellular state.
    The organization of the present paper is as follows. 
    In Sec.\ref{sec:model}, we describe the adopted cell model.
    It consists of a thousand components whose concentrations change through catalytic reaction dynamics.
    Evolution of the network is introduced so that cellular fitness as measured growth rate increases.
    In Sec.\ref{sec:evolution_from_randomly_generated_genotype}, through the numerical evolution of the cellular reaction network, we demonstrate how the dimension reduction emerges through evolution in a thousand-dimensional space.
    By computing the singular values of the inverse Jacobian matrix for the relaxation dynamics of a cellular state, we demonstrate that the largest singular value is separated from all others.
    In the following sections, adaptive evolution to novel or fluctuating environments is discussed in relation to the dimension reduction.
    In Sec.\ref{sec:evolution_from_evolved_genotype}, by taking cells that have evolved to adapt to one environment, we study evolution to a novel environment and how the dimension reduction already shaped by previous evolution provides a constraint on evolution in the novel environment. 
    Interestingly, this constraint can accelerate evolution under novel conditions.
    Sec.\ref{sec:convergence_of_phenotypic_evolutional_pathways} is devoted to explaining phenotypic convergence during evolution in a novel environment.
    In Sec.\ref{sec:evolution_under_fluctuating environment}, evolution in ﬂuctuating environments is studied to show that the dimension reduction is valid under such conditions.
    Sec.\ref{sec:discussion} is devoted to the summary and discussion, where the relevance of the dimension reduction to biology is discussed. 
\section{MATHEMATICAL DESCRIPTION OF DIMENSION REDUCTION}
\label{sec:theory_of_dimension_reduction}
    In this section, we present a general formulation of the response of the phenotypic state $\boldsymbol{x^*}$ upon perturbation.
    It is reformulated from Ref.\cite{Furusawa2018} and is valid not only for the cell model adopted in the present paper, but also generally applicable to phenotypes given by the fixed point of any dynamical systems.
    
    Consider the following $N$-dimensional dynamical system;

    \begin{equation}
        \dot{\boldsymbol{x}}=\boldsymbol{f}(\boldsymbol{g_0}, \boldsymbol{x}), 
        \label{eq:dynamics_theory}
    \end{equation}
    where $\boldsymbol{x}=(x_1, x_2,\dots, x_N)$ is an $N$-dimensional vector of the state variable for the dynamical system, and $\boldsymbol{g_0}$ is a set of parameters characterizing the dynamical system and corresponding to the genotype in our model.
    The phenotype, on the other hand, is given by the fixed point of $\boldsymbol{x}$, which is denoted by $\boldsymbol{x}^*=(x^*_1, x^*_2,\dots, x^*_N)$, which is given by

    \begin{equation}
        \boldsymbol{f}(\boldsymbol{g_0},\boldsymbol{x}^*(\boldsymbol{g_0}))=0.
        \label{eq:fix_theory}
    \end{equation}

    Now consider a slight parameter change, $\boldsymbol{g_0}\rightarrow \boldsymbol{g_0} + \boldsymbol{\delta g}$, due to mutation.
    The fixed point for the modified parameter values is given by

    \begin{equation}
        \boldsymbol{f}(\boldsymbol{g_0}+\boldsymbol{\delta g},\boldsymbol{x}^*(\boldsymbol{g}+\boldsymbol{\delta g}))=0,
    \end{equation}
    from which one gets

    \begin{align}
        \boldsymbol{f}(\boldsymbol{g_0}+\boldsymbol{\delta g},&\boldsymbol{x}^*(\boldsymbol{g_0}+\boldsymbol{\delta g})) \nonumber\\
        &\simeq\boldsymbol{f}(\boldsymbol{g_0},\boldsymbol{x}^*(\boldsymbol{g_0}))+\boldsymbol{R_g}\boldsymbol{\delta g}+\boldsymbol{J}\boldsymbol{\delta x}^*=0,
        \label{eq:fix_delta_theory}
    \end{align}
    where $\boldsymbol{J}=(\partial\boldsymbol{f}/\partial\boldsymbol{x})_{\boldsymbol{x}=\boldsymbol{x}^*(\boldsymbol{g_0})}^{\boldsymbol{g}=\boldsymbol{g_0}}$ is the Jacobian matrix at the fixed point, and $\boldsymbol{R_g}=(\partial\boldsymbol{f}/\partial\boldsymbol{g})_{\boldsymbol{x}=\boldsymbol{x}^*(\boldsymbol{g_0})}^{\boldsymbol{g}=\boldsymbol{g_0}}$ is  the "susceptibility tensor" against parameter change.
    From Eq.(\ref{eq:fix_theory}) and Eq.(\ref{eq:fix_delta_theory}), the change of the fixed point by the parametric change $\boldsymbol{\delta g}$ is given by 

    \begin{equation}
        \boldsymbol{\delta x}^* \simeq -\boldsymbol{L}\boldsymbol{R_g}\boldsymbol{\delta g},
        \label{eq:relation1}
    \end{equation}
    where $\boldsymbol{L}=\boldsymbol{J}^{-1}$ is the inverse Jacobian matrix.
    Moreover, by applying SVD (Singular Value Decomposition) to $\boldsymbol{L}$, it can be decomposed to $\boldsymbol{L}= \boldsymbol{V\Sigma U^{T}}$, where $\boldsymbol{V}=[\boldsymbol{v^{(1)}}, \boldsymbol{v^{(2)}},\dots,\boldsymbol{v^{(N)}}]$ and $\boldsymbol{U}=[\boldsymbol{u^{(1)}}, \boldsymbol{u^{(2)}},\dots,\boldsymbol{u^{(N)}}]$ are orthogonal matrices whose columns are the left and right-singular vectors of $\boldsymbol{L}$, and $\boldsymbol{\Sigma}$ is diagonal matrix whose elements are the singular values $\sigma_i\ (i=1,2,\dots,N)$.
    
    Then, Eq.(\ref{eq:relation1}) can be written in the following form:

    \begin{equation}
        \boldsymbol{\delta x}^*\simeq-\left(\sum_{i=1}^N\sigma_i\boldsymbol{v^{(i)}}\boldsymbol{u^{(i)T}}\right)\boldsymbol{R_g}\boldsymbol{\delta g}.
        \label{eq:relation2}
    \end{equation}

    Note that from the stability of the fixed point, $\sigma_i > 0$ follows. 
    The timescale for relaxation to the fixed point along each mode $v_i$ upon some perturbation is given by $\sigma_i$.
    If the singular values are separated, so that $\sigma_1>\sigma_2,\dots,\sigma_l\gg\sigma_{l+1}>\dots$, then the state change upon perturbation is restricted to the $l$-dimensional subspace spanned by $\boldsymbol{v^{(1)}}, \boldsymbol{v^{(2)}},\dots,$ and $\boldsymbol{v^{(l)}}$:

    \begin{equation}
        \boldsymbol{\delta x}^* \simeq-\left(\sum_{i=1}^{l} \sigma_i \boldsymbol{v^{(i)}}\boldsymbol{u^{(i)T}}\right)\boldsymbol{R}_g\boldsymbol{\delta g}.
        \label{eq:relation_comp}
    \end{equation}
    
    Note that, in some cases, there are some constraints on the dynamics of the model, for example, $\sum_{i=1}^Nx_i=1$ in the model to be described in the next section and adopted in the present paper.
    There also exist trivial null exponents that correspond to the conserved quantities.
    To remove such trivial directions, we compute $\boldsymbol{J}$ in the subspace except in the directions corresponding to the constraints.
\section{MODEL}
\label{sec:model}
    \subsection{Cell model}
    \label{ssub:cell_model}
        To examine the hypothesis that high-dimensional phenotypic changes are restricted to a low-dimensional space as a result of adaptive evolution, we adopted a cell model with a large degree of freedom.
        In this model, changes in cellular state follow genetically determined catalytic reactions.
        We focused on the cellular state of steady growth as a phenotype.
        The model, albeit simplistic, includes the basic, minimum properties of cells, i.e., absorption of nutrients from the environment and their conversion to the components essential for the cell growth, including catalysts for cellular reactions \cite{Furusawa2003}.
        
        In the model, a cell consists of $N$ species of components; thus, the cellular state is represented by the $N$-dimensional vector $\boldsymbol{x}=(x_1,x_2,\dots,x_N)$, where $x_i$ is the concentration of each component.
        We assume that the summation of the concentrations of all components in the cell is constant, which can be set to 1 without loss of generality ($\sum_{i=1}^Nx_i=1$).
        In other words, each concentration $x_i$ is given by the abundance of the $i$th chemical divided by the total abundance of all chemicals.
        This assumption is equivalent to cellular volume increasing in proportion to the total abundance of cellular components.
        The cell model consists of 3 different types of species components: nutrients, transporters, and catalysts.
        Nutrient components exist in the environment and are absorbed into the cell with the aid of cellular transporter chemicals \cite{Furusawa2012AdaptationCriticality}.
        For simplicity, we assumed one transporter per nutrient species.
        The concentrations of nutrients and transporters are given by $\boldsymbol{x_{nut}}=(x_1, x_2, \dots, x_n)$ and $\boldsymbol{x_{tr}}=(x_{n+1}, x_{n+2},\dots,x_{2n}) $\, respectively ($n$ is the number of species of nutrient components).
        The flow rate of each nutrient is given by $Ds_ix_{i+n}$, where $D$ is the parameter of flow rates of nutrients, and $s_i$ is the concentration of the $i$th component in the environment.
        Other components $(k=2n+1, \cdots, N)$ work as catalysts that catalyze $2$-body chemical reactions ( $j+k\rightarrow i+k$ ) in a cell.
        By using the rate equation of catalytic chemical reactions, absorption of nutrients, dilution due to cellular volume growth, and the dynamics of chemical concentrations are given by

        \begin{align}
            \dot{x_i} &= R_i(\boldsymbol{G}, \boldsymbol{x}) + Ds_ix_{i+n} - \mu(\boldsymbol{x}) x_i
            \label{eq:dynamics}\\
            R_i(\boldsymbol{G}, \boldsymbol{x}) &= \sum_{j,k=1}^NG_{ijk}x_jx_k - \sum_{j,k=1}^NG_{jik}x_ix_k
            \label{eq:chemical_reaction}
        \end{align}

        Each term in Eq.(\ref{eq:dynamics}) represents the conversion of components in a cell by catalytic chemical reactions $(j+k\rightarrow i+k)$, absorption of nutrients from the environment, and the dilution effect that accompanies cellular volume growth.
        The cell volume grows according to the absorption of nutrients.
        Hence, the growth rate $\mu(\boldsymbol{x})$ is  given by $\mu(\boldsymbol{x})=\sum_{j=1}^nDs_jx_{j+n}$, which is derived from the hypothesis $\sum_{i=1}^N x_i = 1$.
        $\boldsymbol{G}=\{G_{ijk}\}$ is a 3rd-order tensor corresponding to the genotype of the cell, which takes the value of 1 when a catalytic chemical reaction ($ j+k \rightarrow i+k $) can occur in a cell and takes the value zero otherwise.
        For simplicity, we assumed all reaction coefficients take the same value, 1.
        Unless otherwise noted, we used $N=1000, n=10, D= 0.001\ (=1/N)$ in the present paper.
        
        In the parametric region we adopted in the present paper, cellular states reached a unique, nontrivial fixed point ($\forall i ; x_i^*>0$).
        The concentrations at the fixed point $\{x_i^*\}$ give the phenotype of the cell, which is uniquely determined for a given genotype and environment.
    \subsection{Evolution}
    \label{sub:evolution}
        As the initial genotype before evolution, we generated a catalytic chemical reaction network by randomly putting $\rho N$ reaction paths from each $N$ component and allocating a catalytic component randomly to each chemical reaction path avoiding autocatalytic reactions.
    
        Evolution simulation was carried out in the following procedure.
        At the first generation, we prepared an initial population of $L$ ancestral cells.
        In each generation, from each of the $L$ mother cells, we produced $c$ different mutants and calculated their growth rates $\mu$ with the concentrations of components in the steady state $\boldsymbol{x^*}$.
        This gives the fitness of the cell.
        Then, the top $L$ cells with the highest fitness values were selected to produce offspring for the next generation.
        In the present paper, we used $L=10, c=10$, so that the total cell population was set at $cL=100$.

        Unless otherwise noted, mutations were carried out in the following procedure. 
        At first, we randomly picked a pair of catalytic chemical reactions $(i,j,k), (i',j',k')$, where $(i,j,k)$ represents the catalytic chemical reaction $(j+k\rightarrow i+k)$.
        In the case that $i{\neq}j', i'{\neq}j, j'{\neq}k, j{\neq}k'$, we changed the pair of the reaction paths $(i,j,k), (i',j',k')$ to $(i,j',k), (i',j,k')$ or to $(i',j,k), (i,j',k')$.
        For all the evolution simulations, mutation rate was fixed at 0.04 per component so that 40 paths were changed per generation.
        With this mutation procedure, the total number of incoming and outgoing paths and catalysts, $I_i = \sum_{jk}G_{ijk}$,\ $O_j = \sum_{ik}G_{ijk}$\ and $C_k = \sum_{ij}G_{ijk}$ are conserved.
\section{RESULTS}
    \subsection{Evolution from random network}
    \label{sec:evolution_from_randomly_generated_genotype}
        First, we analyzed cellular evolution with random catalytic chemical reaction networks, following the scheme discussed in Sec.\ref{sub:evolution}.
        In this section, we used the fixed environmental condition, $\boldsymbol{s}^{old}$, through the evolution as
        \begin{equation}
            s_i^{old} = \begin{cases}
                 \ 2/n &(\ i=1,2,\cdots,n/2\ )\\
    
                 \ \ 0 &(\ i=n/2+1,n/2+2,\cdots,n\ ).
            \end{cases}
            \label{eq:env_random}
        \end{equation}
    
\begin{figure}
            \centering
            \includegraphics[clip,width=0.9\linewidth]{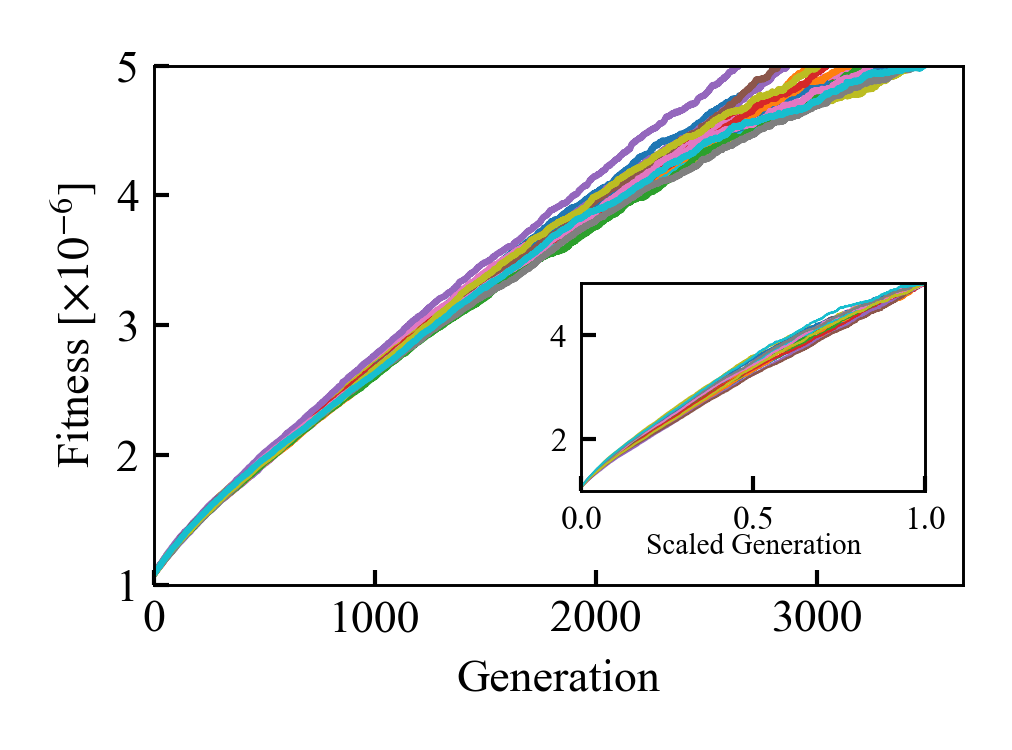}
            \caption{Evolutionary courses of maximum fitness in the population starting from cells with randomly chosen reaction networks. 
            Different colors correspond to different runs with different random mutations, starting from the same ancestral cell in the same environment. 
            The speed of the fitness increase differed in each run. 
            Inset: we used the “scaled generation” for the horizontal axis to normalize a given fitness maximum ( here, $5\times 10^{-5}$ ) to 1. 
            In the figures below, this rescaling is often adopted to compare the results of different runs.}
            \label{FIG:evo_from_random_1}
        \end{figure}

        Through evolution, maximum fitness in the population monotonically increased (Fig.\ref{FIG:evo_from_random_1}).
        In Fig.\ref{FIG:evo_from_random_1}, the top fitness values in the population are plotted across different strains (i.e., for different runs of the evolution simulation) in different colors. 
        For all runs with different random mutations, fitness increased to a sufficiently high level, and the time course of the increases, after rescaling the generations, was approximately the same.
        
\begin{figure*}
            \centering
            \includegraphics[clip,width=0.9\linewidth]{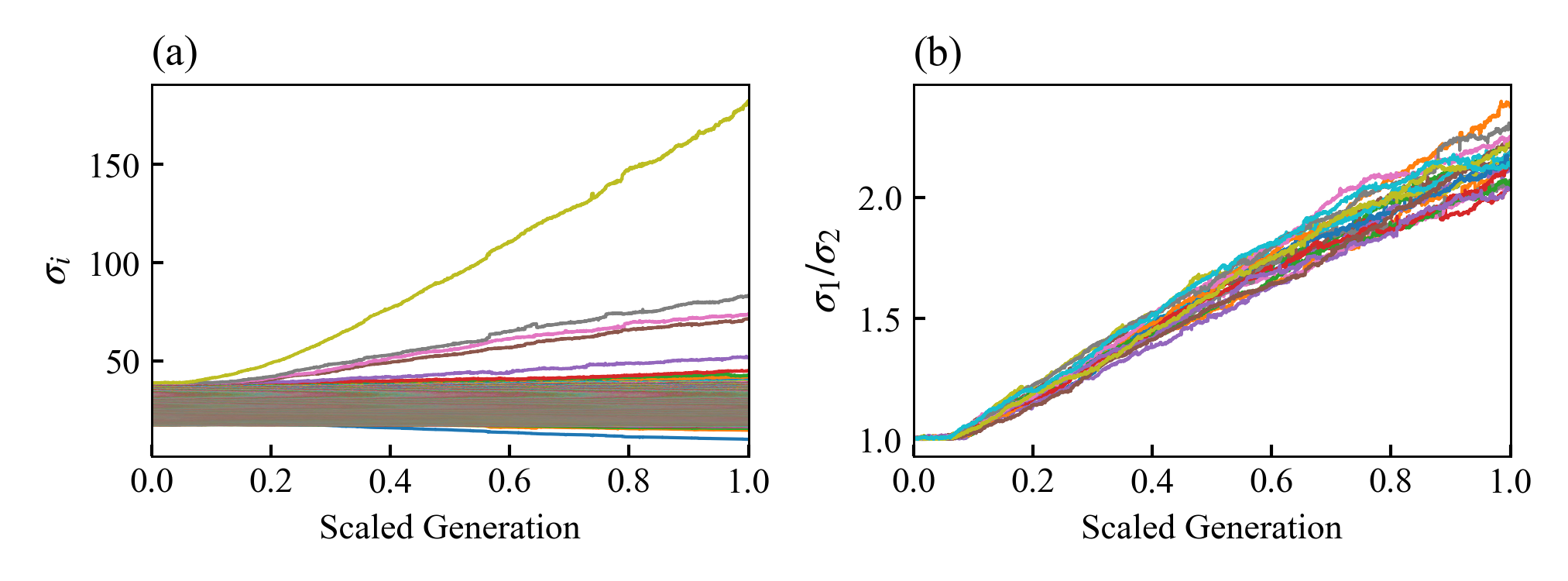}
            \caption{
            (a)\ Evolutionary changes of singular values $\{\sigma_i\}$ of the inverse Jacobian matrix $\boldsymbol{L}$ for the phenotype $\boldsymbol{x}^*$. 
            They are calculated for the fittest cells in the population every generation.
            (b)\ Evolutionary changes of the ratio of the first to the second largest singular values $\sigma_1/\sigma_2$ for different strains.
            Both (a) and (b) are plotted against the scaled generation.}
            \label{FIG:evo_from_random_2}
        \end{figure*}

        Next, we computed the Jacobian matrix $\boldsymbol{J}$ at the fixed point $\boldsymbol{x^*}$ for the reaction dynamics at each generation.
        We then obtained the singular values $\{\sigma_i\}\ (\sigma_1>\sigma_2>\dots>\sigma_{N-1})$ of its inverse matrix $\boldsymbol{L}$ (see also Eq.\ref{eq:relation_comp}) and the corresponding left-singular vectors $\{\boldsymbol{v^{(i)}}\}$.
         As shown in Fig.\ref{FIG:evo_from_random_2}(a), the largest singular value $\sigma_1$ was separated from the other singular values through evolution. 
        This trend was common across all evolutionary runs.
        The ratio of the first to the second largest singular values $\sigma_1/\sigma_2$ increased through evolution (Fig.\ref{FIG:evo_from_random_2}(b)).
        This suggests that the relaxation dynamics of phenotypes are highly constrained along $\boldsymbol{v^{(1)}}$, the left-singular vector for the largest singular value $\sigma_1$.
        
\begin{figure*}
            \centering
            \includegraphics[clip,width=0.9\linewidth]{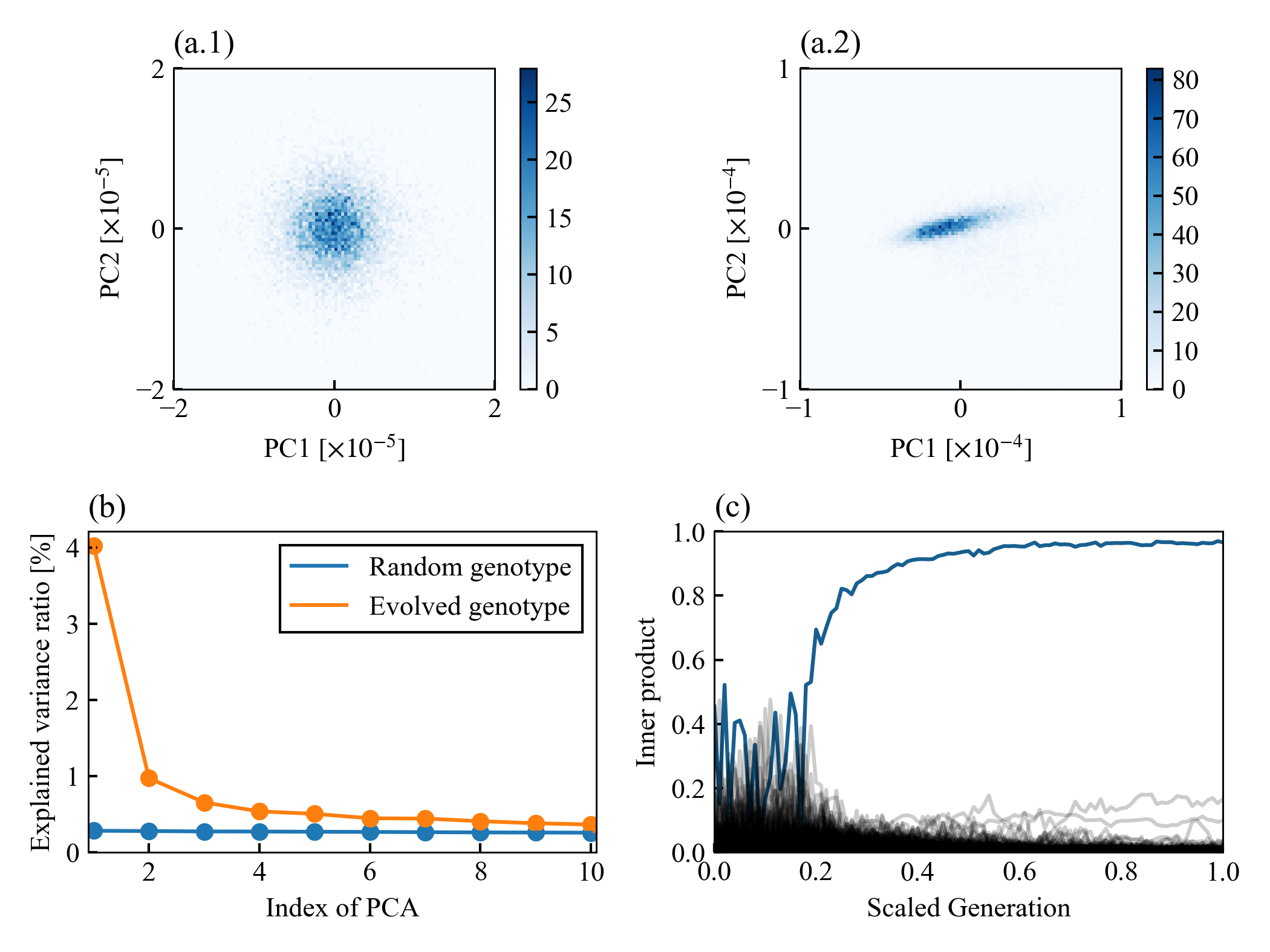}
            \caption{(a)\ Distribution of phenotypic changes due to genetic mutations in the PC (principal component) plane at the start (a.1) and end of evolution (a.2). 
            These PCs were calculated from the phenotypes $\boldsymbol{x^*}$ of $10^5$ mutants.
            (b)\ Explained variance ratio of the PC modes of phenotypic changes due to genetic mutation. 
            The blue line corresponds to the data at the start of evolution and the orange to that at the end of evolution.
            (c)\ Inner products between the 1st PC mode of the phenotypic changes due to mutation at each generation and the left-singular vectors $\boldsymbol{v^{(i)}}$ of the inverse Jacobian matrix $\boldsymbol{L}$.
            The blue line shows the inner product with  $\boldsymbol{v^{(1)}}$, the left-singular vector of the largest singular value, and orange lines depict those with the other singular vectors, plotted against the rescaled generation.}
            \label{FIG:evo_from_random_3}
        \end{figure*}

        Next, we analyzed the distribution of phenotypic changes caused by genetic mutation as follows:
        At each generation, we picked the fittest cell in the population and produced many thousands of mutant cells, which have slightly modified chemical reaction networks from those of the original.
        For each of these mutant cells, the phenotypes $\boldsymbol{x}^*$ were computed.
        Then, each of the high-dimensional phenotypic changes due the mutation was obtained.
        To visualize these changes in the $N$-dimensional space, we used PCA (principal component analysis).
        In Fig.\ref{FIG:evo_from_random_3}(a), we plotted the phenotypic changes using the 1st and 2nd PC modes, $\boldsymbol{p^{(1)}_{old}}$ and $\boldsymbol{p^{(2)}_{old}}$, respectively.
        At the start of evolution, the phenotypic changes of the mutant cells were uniformly distributed in the PC plane (Fig.\ref{FIG:evo_from_random_3}(a.1)).
        As evolution progressed, the distribution of phenotypic changes was largely biased along $\boldsymbol{p^{(1)}_{old}}$ (Fig.\ref{FIG:evo_from_random_3}(a.2)).
        In fact, at the end of the evolution, more than $4\%$ of the phenotypic changes due to mutation were explained by the 1st PC (Fig.\ref{FIG:evo_from_random_3}(b)), though less than $0.3\%$ were explained at the start of the evolution.
        This indicates that phenotypic changes due to mutation were constrained to a lower-dimensional space as the evolution progressed.
        
        
        Next, we studied how the low-dimensional constraint in relaxation dynamics and the low-dimensional response to genetic mutations are correlated.
        As an indicator of the similarity between relaxation dynamics of the phenotype and its response to genetic mutations, we calculated the absolute inner products of $\boldsymbol{p^{(1)}_{old}}$ with $\boldsymbol{v^{(i)}}(i=1,2,\dots,N-1)$ for the fittest cell in each generation (Fig.\ref{FIG:evo_from_random_3}(c)).
        In Fig.\ref{FIG:evo_from_random_3}(c), the inner product with $\boldsymbol{v^{(1)}}$ is depicted by the blue line and the others by the thin black line.
        
        \begin{figure*}
            \centering
            \includegraphics[clip,width=0.9\linewidth]{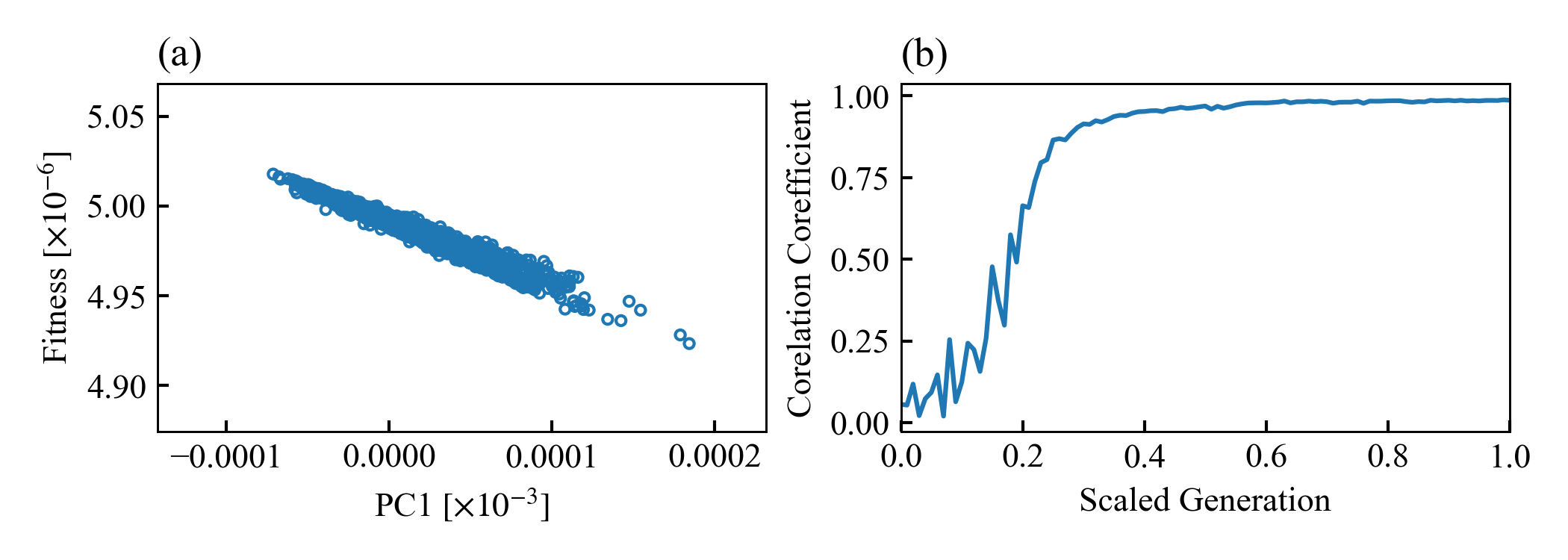}
            \caption{(a)\ Dependency of fitness on PC1 values of the phenotype of mutant cells computed across 10000 mutants after the last generation of evolution.
            (b)\ Correlation between fitness and the PC1 value of the phenotype of mutant cells computed every 10 generations, as above, and plotted against the scaled generation.}
            \label{FIG:evo_from_random_4}
        \end{figure*}
        The former took on a much higher value as the evolution progressed.
        This result shows that phenotypic changes caused by genetic mutation were restricted to the one-dimensional subspace spanned by $\boldsymbol{v^{(1)}}$, i.e., the direction of the slowest relaxation mode, after adaptive evolution had progressed.
        
        Thus, our results demonstrate that a low-dimensional constraint in the phenotypic space emerged through adaptive evolution.
        The next question concerns the biological meaning of this low-dimensional structure.
        We then plotted the dependency of fitness on PC1 value $p^{(1)}_{old}=(\boldsymbol{p^{(1)}_{old}}\cdot \boldsymbol{x^*})$.
        As shown in Fig.\ref{FIG:evo_from_random_4}(a), $p^{(1)}_{old}$ was highly correlated with fitness after evolution.
        Indeed, the correlation between $p_1$ and fitness increased through evolution (Fig.\ref{FIG:evo_from_random_4}(b)).
        This result indicates that the one-dimensional constraint in the phenotypic space both in relaxation dynamics and in response to genetic mutation was acquired through adaptive evolution, which embeds fitness into the most variable direction in the phenotypic subspace.
    \subsection{Evolution in novel environment}
    \label{sec:evolution_from_evolved_genotype}
        In the previous section, we analyzed evolution of the cells with randomly generated chemical reaction networks.
        In nature, however, evolution occurs for organisms that have already evolved under earlier environmental conditions.
        Hence, in this section, we use those already evolved in one environmental condition and study their evolution in a new environment.
        For this, we took cells that had evolved in the fixed environment $\boldsymbol{s}^{old}$ (given by Eq.(\ref{eq:env_evo})) from a random network, and then put them into the following new fixed environment $\boldsymbol{s}^{new}$.
        
        \begin{equation}
            s_i^{new} = \begin{cases}
                 \ \ 0 &(\ i=1,2,\cdots,n/2\ )\\
    
                 \ 2/n &(\ i=n/2+1,n/2+2,\cdots,n\ )
            \end{cases}
            \label{eq:env_evo}
        \end{equation}
    
\begin{figure*}
            \centering
            \includegraphics[clip, width=0.9\linewidth]{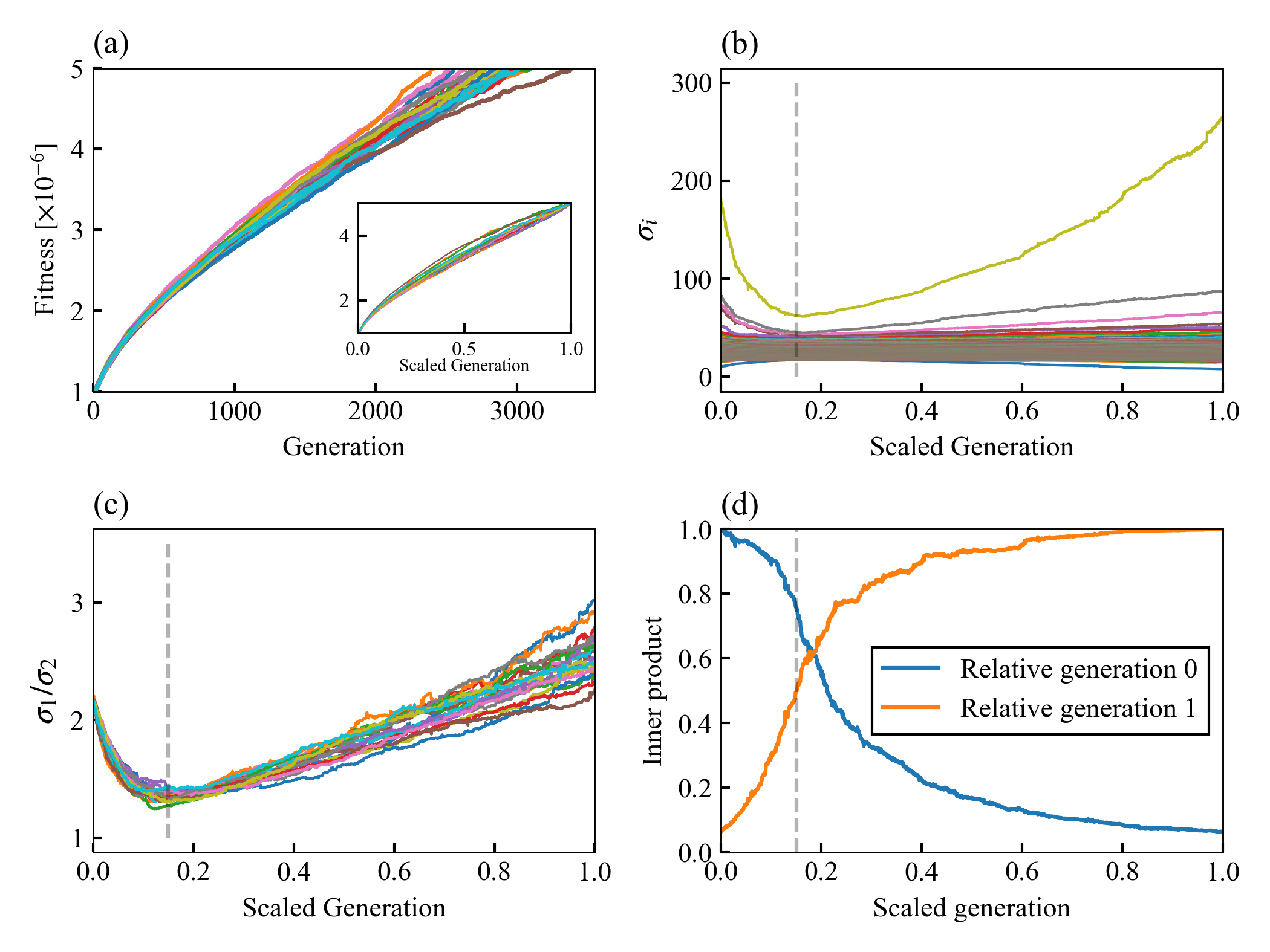}
            \caption{(a)\ Evolutionary courses of maximum fitness in a population evolving in a novel environment starting from cells evolved in another environment. 
            The speed of fitness increase differed in each run. 
            Hence, in the embedded figure, we used the 'scaled generation' for the horizontal axis to normalize a given fitness maximum (here $5\times 10^{-5}$ ) to 1.
            Different colors correspond to different runs with different random mutations starting from the same ancestral cell and environmental conditions.
            (b)\ Evolutionary changes of all the singular values $\sigma_i$.
            (c)\ Evolutionary changes to the ratio of the first to the second singular values.
            (d)\ Evolutionary changes to the inner product between $v_1$ at each generation and that at the start of evolution (blue line) or end of evolution (orange line). 
            All of (a)-(d) are plotted against the scaled generation.}
            \label{FIG:evo_from_evo}
        \end{figure*}

        Again, maximum fitness in the population monotonically increased through evolution of all strains following approximately the same course (Fig.\ref{FIG:evo_from_evo}(a)).
        We then computed the distribution of $\{\sigma_i\}$, the singular values of $\boldsymbol{L}$.
        In this case, the largest singular value $\sigma_1$ remained separated from the other singular values through evolution (Fig.\ref{FIG:evo_from_evo}(b)), whereas $\sigma_1/\sigma_2$ first decreased, and then increased later. 
        Fig.\ref{FIG:evo_from_evo}(c) shows a minimum value approximately at the scaled generation $0.15$.
        
        Next, we computed how the most changeable direction of the phenotype with $\boldsymbol{v^{(1)}}$ changes through evolution to a new environment.
        In each generation, we calculated the inner products between the vector and that either at the start or end of the evolution (Fig.\ref{FIG:evo_from_evo}(d)).
        As shown, the former decreased from $1$ to $0$, whereas the latter increased from $0$ to $1$.
        The latter inner product exceeded the former at the same relative generation (0.15) when $\sigma_1/\sigma_2$ took the minimum value.
        
        The results in the present section imply that adaptation to a novel environment first progressed within the one-dimensional phenotypic constraint imposed by the previous evolution, and that this one-dimensional subspace was later redirected to match the novel environment.
    
    \subsection{Convergence of phenotypic evolutionary pathways}
    \label{sec:convergence_of_phenotypic_evolutional_pathways}
\begin{figure*}
            \centering
            \includegraphics[clip,width=0.9\linewidth]{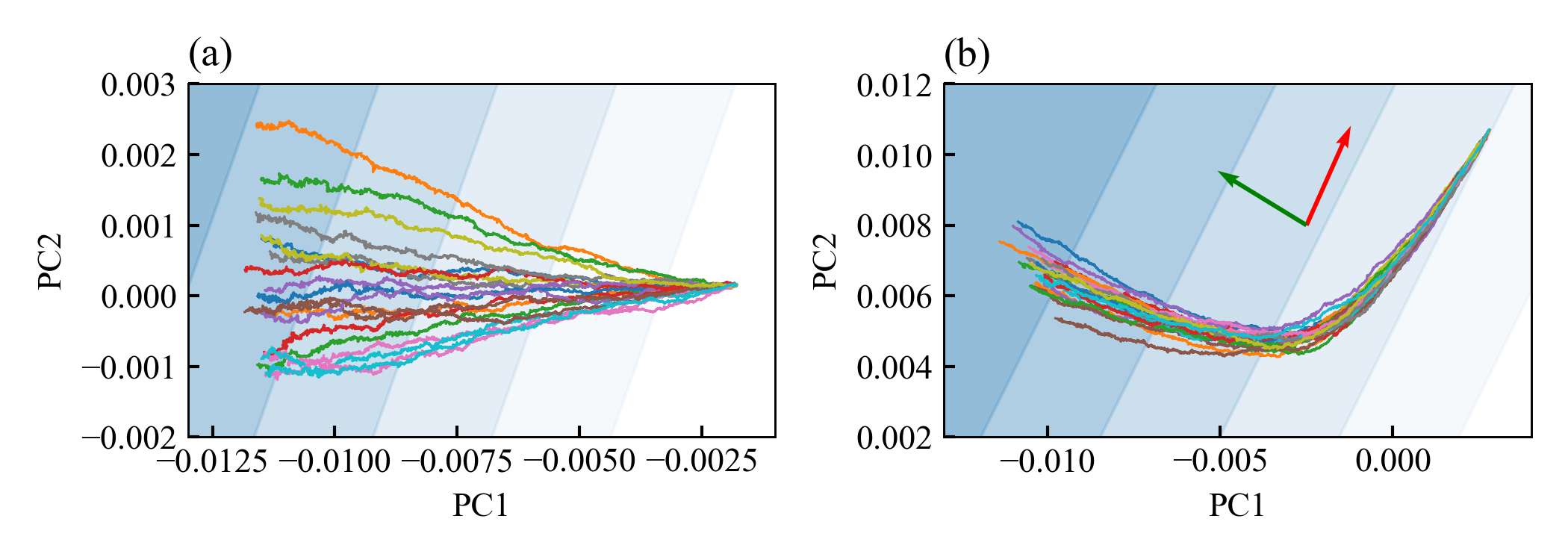}
            \caption{Evolutionary pathways plotted in the plane of PC1 and PC2: (a)\ evolution under the environment condition $\boldsymbol{s^{old}}$ starting from a random network and (b)\ evolution under the condition $\boldsymbol{s^{new}}$ for genotypes already evolved under the condition $\boldsymbol{s^{old}}$.
            The red vector in (b) is the projection of $v_1$, the left-singular vector of the largest singular value of $\boldsymbol{L}$ at the start of the evolution, and the blue one is that at the end of the evolution.
            The gradation in both figures represents the fitness value, with darker color corresponding to higher value as plotted in the PC plane.
            The PC planes are calculated according to the phenotypes of the fittest cells in the population every 0.001 relative generations for the 20 different strains.}
            \label{FIG:pathway}
        \end{figure*}

        In this section, we will investigate evolutionary pathways in the phenotypic space from the phenotype adapted to the old environment.
        We observed that evolutionary pathways in phenotypic space converged as a result of adaptive evolution (Fig.\ref{FIG:pathway}(b)).
        In the figure, the evolutionary pathways of 20 different strains from the same ancestral cell evolved under the same environmental conditions $\boldsymbol{s}^{old}$ are plotted in the plane with the 1st and the 2nd PCs, where the PCs were computed over a set of phenotypes of the 20 strains over 1000 equally divided generations from the start to end of each evolution.
        The gradation in the background of the figure represents the fitness values shown as a function of the PC plane.
        
        Different strains from the same ancestral cell similarly climbed up the fitness landscape due to strong selection during adaptive evolution.
        However, in the neutral direction, in which fitness does not change, the phenotypes of different strains could move in different directions due to random mutations. 
        Indeed, this was the case for the original evolution from the random network (Fig.\ref{FIG:pathway}(a)).
        In contrast, the evolutionary pathways from the cells which had evolved in the old environment $\boldsymbol{s}^{old}$ followed the same path, not only in the direction along which fitness changes but also in the direction along which it does not (Fig. \ref{FIG:pathway}(b)).
    
        The difference between the above two evolutionary pathways originates in how the genetic mutations are mapped to the phenotypic changes.
        For cells with randomly chosen genotypes, phenotypic changes caused by mutations were uniformly distributed in all directions across the phenotypic space.
        Thus, evolutionary pathways diverged in the directions along which fitness does not change.
        On the other hand, for cells evolved in a given environment, the phenotypic changes induced by mutations were restricted to a low-dimensional subspace.
        Therefore, phenotypic changes in later evolutionary cycles were highly biased within this subspace. 
        Thus, the evolutionary pathways in the phenotypic space were restricted to this biased direction.
        
        Thus far, we have shown the convergence of evolutionary pathways of the phenotypes that emerged as a result of low-dimensional constraints on phenotypic changes already realized in earlier evolution under different conditions. 
        One might wonder whether such constraints hinder or foster evolution to a new environment.
    
\begin{figure}
            \centering
            \includegraphics[clip,width=0.9\linewidth]{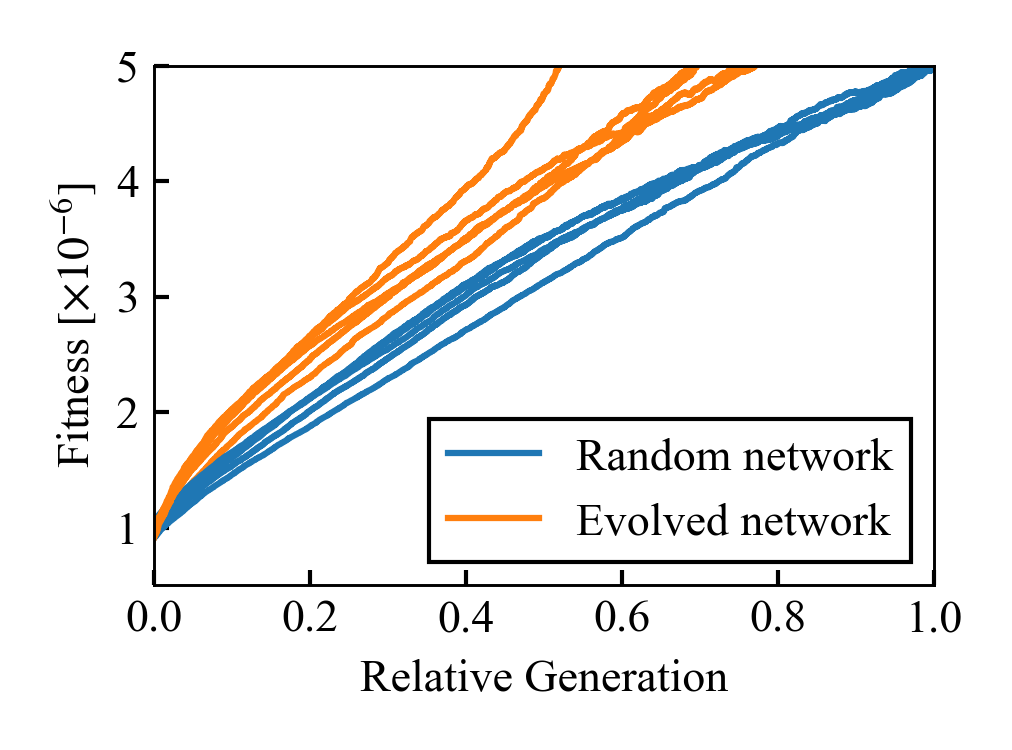}
            \caption{ Fitness changes through the evolution from cells with randomly generated networks  (blue line) and from the cells that have already evolved to an earlier, different environmental condition. }
            \label{FIG:speed_evolution}
        \end{figure}

        To examine these possibilities, we computed evolution speeds with and without such constraints. 
        In Fig.\ref{FIG:speed_evolution}, evolutionary courses of fitness are plotted: one for cells evolved under a previous environmental condition and the other for cells evolved based on a random network.
        The generations needed to reach a given growth rate ($\mu^*=5\times 10^{-6}$) were $20\%$ shorter for the former as compared with the latter, even though initial growth rates were the same.
    
        As shown in Sec.\ref{sec:evolution_from_evolved_genotype}, cells took advantage of the already evolved low-dimensional structure and modified it rather than destroying the structure in order to adapt to a novel environment.
        These results suggest that the low-dimensional phenotypic constraint hastens evolution to a novel environment.
    \subsection{Evolution in fluctuating environment}
    \label{sec:evolution_under_fluctuating environment}
        In the last subsection, we studied adaptation to changes from one environment to another.
        In nature, environmental conditions can continuously change over generations.
        Thus, the next question to be addressed is whether such a one-dimensional phenotypic constraint is also generated through evolution in a fluctuating environment, where environmental conditions repeatedly change after some number of generations.
        To be specific, nutrient conditions were randomly changed every 10 generations to one of 5 different conditions $\boldsymbol{s}^{(i)} (i=1,2,\dots,5)$ as given by
    
        \begin{equation}
            s_j^{(i)} = \begin{cases}
                 (1-\epsilon)/2 &(\ j=2i-1,\ 2i\ )\\
                 \epsilon/(n-2) &(\ \rm{the\ others}\ )
            \end{cases} 
            \label{eq:env_fluc}
        \end{equation}  
        with $\epsilon=0.2$ and $n=10$.
        Note that the following results are qualitatively consistent with the intervals for changing environmental conditions.
        Herein, we term this changing environmental condition $E_s$, whereas each of the 5 nutrient conditions are termed $E_1, E_2, \dots, E_5$, respectively.
        \begin{figure*}
            \centering
            \includegraphics[clip,width=0.9\linewidth]{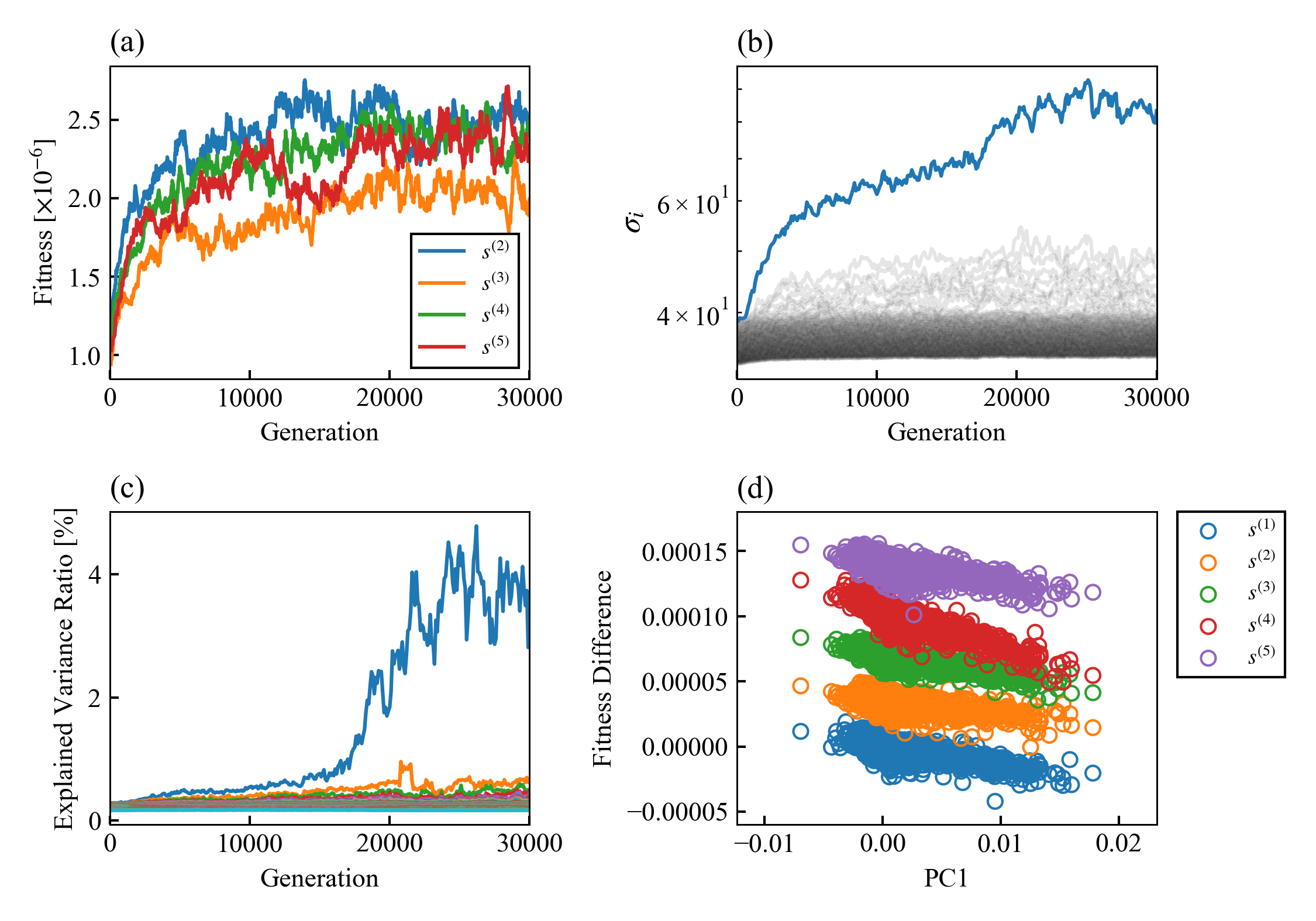}
            \caption{(a)\ Evolutionary changes to fitness value under 5 environmental conditions for the network with the highest fitness under a fluctuating environment.
            (b)\ Evolutionary changes to the singular values $\sigma_i$ of the inverse Jacobian matrix $\boldsymbol{L}$.
            (c)\ Evolutionary changes to the explained variance ratio of PCs calculated from the phenotypic changes in many thousands of mutants in each generation.
            (d)\ Relation between the 1st PC mode of the phenotypic changes caused by mutations and fitness in different environments.
            The vertical axis, fitness difference, represents the fitness at each environmental condition subtracted from the average.
            Those in all 5 environments showed similar trends.
            }
            \label{FIG:evo_multi_1}
        \end{figure*}
        
        By carrying out numerical evolution, we confirmed that the growth rate of the fittest genotype increased in all 5 environmental conditions, as shown in Fig.\ref{FIG:evo_multi_1}(a).
        We then computed the singular values $\{\sigma_i\}$ of the inverse Jacobian matrix $\boldsymbol{L}$ in the same way as in the previous subsections.
        Again, only one largest singular value $\sigma_1$ was separated from the others in spite of the diversity of environmental conditions (Fig.\ref{FIG:evo_multi_1}(b)).
        Phenotypic changes due to genetic mutation were again restricted to the one-dimensional subspace spanned by $\boldsymbol{v^{(1)}}$, the left-singular vector corresponding to $\sigma_1$ (Fig.\ref{FIG:evo_multi_1}(c)).
        For convenience, we label this 1st PC mode of the phenotypic changes due to genetic mutation in cells evolved under $E_s$ as $\boldsymbol{p^{(1)}_s}$, whereas each of the 1st PC modes computed from the network evolved separately under each $\boldsymbol{E_i}$ condition is denoted  as $\boldsymbol{p^{(1)}_i} (i=1,2,\dots,5)$.
        
        We then examined whether and how $\boldsymbol{p^{(1)}_s}$ represents fitness in all 5 conditions.
        As shown in Fig.\ref{FIG:evo_multi_1} (d), the inner product $p^{(1)}_s=(\boldsymbol{p^{(1)}_s}\cdot \boldsymbol{x^*})$ correlated well with fitness in all 5 conditions.
        This suggests that $\boldsymbol{p^{(1)}_s}$ involves all modes evolved in each environmental condition $E_i\ (i=1,2,\dots,5)$.
        To examine how $\boldsymbol{p^{(1)}_s}$ is related to phenotypic constraint of the networks evolved in each condition $E_i (i=1,2,\dots,5)$, we computed the similarity of $\boldsymbol{p^{(1)}_s}$ with each of $\boldsymbol{p^{(1)}_i} (i=1,2\dots,5)$, as given by the absolute inner products $\boldsymbol{p^{(1)}_s}\cdot\boldsymbol{p^{(1)}_i}$.
        
        Firstly, the inner products $|(\boldsymbol{p^{(1)}_i}\cdot\boldsymbol{p^{(1)}_j})| (i\neq j=1,2,\dots,5)$ were smaller than 0.006; i.e., $\boldsymbol{p^{(1)}_i}$s were almost orthogonal to each other.
        In contrast, $a_i\equiv(\boldsymbol{p^{(1)}_s}\cdot\boldsymbol{p^{(1)}_i}) (i=1,2,\dots,5)$ were 0.23, 0.076, 0.11, 0.20, and 0.12, respectively.
        These values were much larger than $|(\boldsymbol{p^{(1)}_i}\cdot\boldsymbol{p^{(1)}_j})|$.
        Since $\boldsymbol{p^{(1)}_i}$s are orthogonal to each other, $\boldsymbol{p^{(1)}_s}$ is represented by the linear combination $\boldsymbol{p^{(1)}_s}\simeq\sum_{i=1}^5a_i\boldsymbol{p^{(1)}_i}+(others)$.
        %with $a_i=(\boldsymbol{p^s}\cdot\boldsymbol{p_i})$.
        If $\boldsymbol{p^{(1)}_s}$ were completely represented by the linear combination of $\boldsymbol{p^{(1)}_i}$s with the same weight and no other contributions, $\boldsymbol{p^{(1)}_s}=\frac{1}{\sqrt{5}}\sum_{i=1}^5\boldsymbol{p^{(1)}_i}$ would hold.
        The observed inner products $a_i$ were smaller than this value $1/\sqrt{5}\sim0.42$, but were substantially large.
        This indicates that the most variable direction in phenotypic changes acquired through changing environmental conditions mainly reflected each of the variable modes acquired for each environmental condition, but also included some others which may be necessary for adaptation to changing environments.
\section{Discussion}
\label{sec:discussion}
    In the present paper, we studied the evolution of a catalytic reaction network to higher fitness as measured by growth rate to confirm that the phenotypic change in a high-dimensional space of chemical concentrations is restricted mainly within a one-dimensional subspace after evolution. 
    This was demonstrated by the observation that one singular value of the inverse Jacobian matrix at the steady growth state is significantly larger than others. 
    Along the direction of the left-singular vector corresponding to this value, the relaxation is slowest.
    Thus, phenotypic change is constrained in this direction.
    The drastic dimension reduction from high-dimensional phenotypic changes to this one-dimensional subspace emerges after evolution. 
    Furthermore, this one-dimensional direction agrees with the direction in dominant phenotypic changes due to mutation from the fitted network. 
    The most variable direction in the reaction dynamics agrees with the most variable direction due to genetic changes.
    In addition, along this one-dimensional subspace, the fitness gradient is highest, implying that this left-singular vector represents fitness. 

    Next, it was shown that this one-dimensional phenotypic space evolved under a given environmental condition was used for later adaptation to a novel environment. 
    Evolution is accelerated by this dimension reduction. 
    Further, this reduction to one-dimensional phenotypic space emerges even under fluctuating environments.

    Dimension reduction from high-dimensional phenotypic states has gathered much attention recently in the studying biological systems and in protein dynamics \cite{Kaneko2015,Carroll2013,Pancaldi2010Meta-analysisYeast, Keren2013, Tlusty2017PhysicalProteins, Schreier2017ExploratoryNetworks, Frentz2015StronglyCommunities, Daniels2008SloppinessBiology}.
    Changes in the concentrations of most mRNA and protein species in response to environmental stress are suggested to be mainly restricted to a one-dimensional subspace.
    Thus, the present results of evolved catalytic reaction networks are consistent with previous experimental observations. 
    Notably, such a reduction to a one-dimensional subspace from a high-dimensional phenotypic space is explained by the separation of a single singular value of the inverse Jacobian matrix around the steady state from all others.
    Changes upon perturbation are mainly constrained along the left-singular vector of this outlier singular value, and accordingly, the one-dimensional constraint is derived as a result of evolution.

    Can one then directly confirm this separation of one singular value in experiments? Upon perturbation, cellular states relax to the original steady state. 
    The slowest relaxation mode is given by the left-singular vector for the separated singular value of the Jacobian matrix. 
    For example, by means of transcriptome analysis, Braun measured the temporal course of the concentrations of all mRNA species upon environmental stress \cite{Stolovicki2011a}.
    Interestingly, the changes were highly correlated across all species, suggesting constraint of relaxation dynamics to a low- (one-) dimensional subspace.

    An alternative possibility for experimental confirmation of singular value separation would be the use of the concentration fluctuations of chemicals inevitable in a cell due to the small number of molecules within it \cite{Elowitz2002StochasticCell.,Furusawa2005UbiquityDynamics,Bar-Even2006NoiseAbundance,Sato2003}. 
    If the fluctuations are mainly governed by the slowest mode, i.e., that corresponding to the largest singular value of the inverse Jacobian matrix, then one possible strategy is to measure the temporal fluctuations in chemical concentrations (gene expression) and their correlation matrix. 
    Even though it may be difficult to measure such fluctuations for many components at a single-cell level, the information could be collected from a population \cite{Brenner2015}.

    It is interesting to note that one-dimensional constraint in concentration changes is observed even during evolution under fluctuating environmental conditions. 
    Even though the fittest phenotype and the corresponding one-dimensional subspace is different for each environmental condition, the evolved state under the fluctuating environment still has only one distinct singular value, and the cellular state is dominantly changeable along its left-singular vector.
    This left-singular vector (manifold) keeps some overlap with that shaped for each of the environments, but is not just the combination of each.
    The one-dimensional phenotypic space retaining the information of each environmental condition as well as that needed for adaptation to environmental conditions that switch over generations is shaped through the evolution.

    Why was only a one-dimensional structure acquired in evolution of the present cell model? 
    In the present study, we used cell growth rate as an indicator of fitness, even though we changed the environmental conditions. 
    In reality, cells often need to evolve to satisfy other conditions, such as survival in poor nutrient conditions or resistance to antibiotics, in addition to the pressure for higher growth. 
    To confirm that the reduction to a one-dimension subspace is valid even for different indicators of fitness in evolution, we carried out evolution simulations of fitness that were not correlated with growth rate (See Supplemental Materials).
    To summarize the result of these simulations, a one-dimensional constraint with separation of a single singular value always evolved.
    The specific choice of the fitness function in the simulation is not important to the dimension reduction for phenotypic changes.
 
    In reality, however, phenotypic changes in bacteria are not necessarily restricted to a one-dimensional subspace.
    For example, major phenotypic changes in {\sl E. Coli} under applications across a variety of antibiotics were recently measured by transcriptome analysis.
    Even though a dimension reduction was observed, the transcriptome changes were located in a subspace of more than one-dimension, possibly in an approximately 8-dimensional subspace \cite{Suzuki2014b}.
 
    One possible reason that the present model produces a lower-dimensional subspace than does real-life omics analysis may be the lack of an explicit tradeoff, an important topic in the study of evolution \cite{Kashtan2005, Shoval2012EvolutionarySpace, Tendler2015EvolutionaryShells}.
    As shown in Fig.\ref{FIG:evo_multi_1} (d), fitness changes across the different environmental conditions were highly correlated. 
    This positive correlation indicates that there existed no tradeoff among the fitness pressures under the different conditions. 
    Thus, whether the dimensionality of constrained phenotypic space depends on environmental conditions including explicit tradeoffs is an important question.

    As shown in Sec.\ref{sec:convergence_of_phenotypic_evolutional_pathways}, once a low-dimensional structure is formed as a result of evolution, evolutionary pathways in the phenotypic space follow similar trajectories in different populations.
    Indeed, this convergence of evolutionary pathways was previously observed in an evolution experiment with bacteria \cite{Horinouchi2015b}.
    According to our results, this convergence is explained as follows.
    Evolutionary pathways in the phenotypic space are determined by selection according to fitness and drift due to genetic mutation.
    Under a low-dimensional phenotypic constraint, phenotypic changes due to genetic drift are restricted to the low-dimensional subspace. 
    In this way, evolutionary pathways are strongly biased in this direction as shown in Fig.\ref{FIG:pathway}. Hence, possible evolutionary pathways in the phenotypic space are highly restricted within the low-dimensional subspace. This may shed new light on the long-lasting question of “necessity or chance” in evolution (or “replaying the tape of life” as described by Gould \cite{GouldWonderfulHistory}); even though genetic evolution could be random, phenotypic evolution is rather deterministic and constrained.

    Whether or not the low-dimensional constraint imposed by evolution is beneficial remains unclear.
    One benefit we observed is the acceleration of adaptation to novel environments.
    As shown in Fig.\ref{FIG:speed_evolution}, the generations needed to adapt to a novel environment were fewer for the network previously evolved in an earlier different environment than those for the random network. 
    As discussed above, the evolved low-dimensional structure was reused to aid adaptation rather than discarded. Searching for fitter states within the restricted subspace would be generally faster than random searching over a higher-dimensional space. 
    Hence, evolution shapes the low-dimensional structure in the phenotypic space, which accelerates further evolution in turn.
    Of course, this acceleration is only possible if fitter states are accessible in the low-dimensional structure. 
    In this sense, if explicit trade-offs exist between the fitted states for the old and new environmental conditions, acceleration may not occur.
    This issue should be explored in the future.
  
    Our description of the dimension reduction adopts linearization around the steady growth state, which assumes small changes upon perturbation. 
    Although it has been suggested that such a linear regime is expanded as a result of increased robustness through evolution \cite{Furusawa2018, Kaneko2007}, it will be important to examine the ranges at which linearization is valid and also to explore extension of dimension reduction to a nonlinear regime.

    In summary, we have formulated the dimension reduction in biological systems in terms of dynamical systems and confirmed it by simulation of evolution in a cell model with reaction networks of large numbers of chemical species.
    The low-dimensional constraint on phenotypic changes leads to further deterministic phenotypic evolution, which also allows for the acceleration of adaptation to novel environments. 
    These results lay the groundwork for establishing a theory of the constraint and predictability of phenotypic evolution.

\section*{Acknowledgments}
The authors would like to thank Chikara Furusawa, QianYuan Tang, Omri Barak, Tetsuhiro Hatakeyama, and Atsushi Kamimura for stimulating discussions.

This research was partially supported by a Grant-in-Aid for Scientific Research (S) (15H05746) from the Japanese Society for the Promotion of Science (JSPS) and a Grant-in-Aid for Scientific Research on Innovative Areas (17H06386) from the Ministry of Education, Culture, Sports, Science and Technology (MEXT) of Japan.

\appendix
\section{APPENDIX: EVOLUTION OF FITNESS UNCORRELATED TO GROWTH RATE}
\label{app:evolution_with_fitness_uncorrelated to growth rate}
\begin{figure*}
        \centering
        \includegraphics[clip, width=0.9\linewidth]{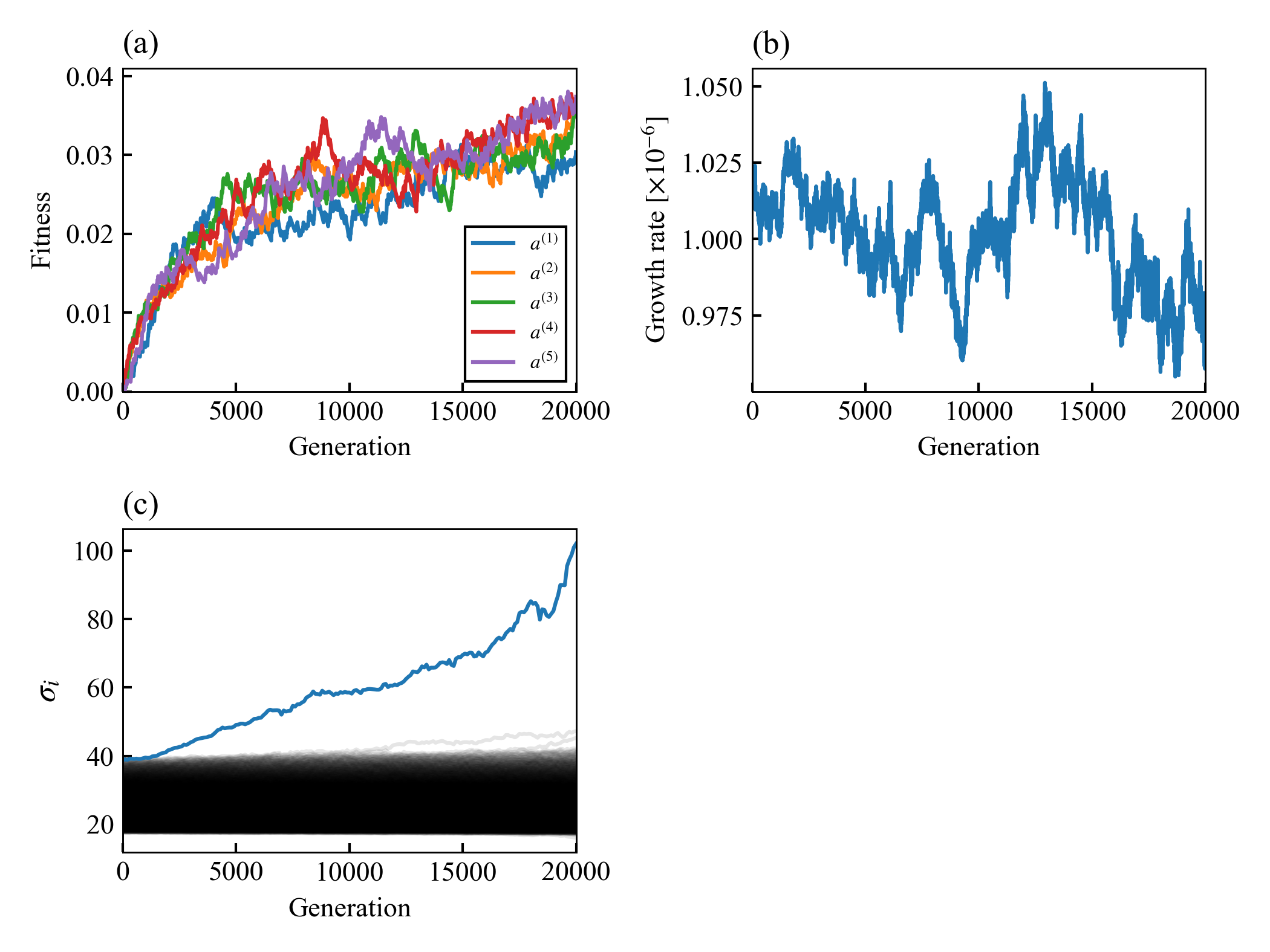}
        \caption{Evolutionary changes to maximum fitness in a population (a), the growth rate (b), and the singular values (c) of the fittest cells using the fitness function described in Eq.\ref{eq:fitness_function_non-growth}. 
        In (a), the results of 5 runs are overlaid, whereas (b) and (c) show the results of one run.}
        \label{FIG:Non-growth}
    \end{figure*}

    One-dimensional constraint of phenotypic changes in the main text was obtained using growth rate as the indicator of evolutionary fitness.
    To confirm the generality of this result for different choices of indicator, we adopted the fitness function $f(\boldsymbol{x^*};\boldsymbol{w^{(i)}})$;
    
    \begin{equation}
        f(\boldsymbol{x^*};\boldsymbol{a^{(i)}}) = \sum_{j=1}^{k-2n}w_{j}^{(i)}x_{j+2n}^*
        \label{eq:fitness_function_non-growth}
    \end{equation}
    where $\boldsymbol{w^{(i)}}=(w^{(i)}_1, w^{(i)}_2, \dots, w^{(i)}_{k-2n})$ is a vector whose elements are generated by uniform distribution between -1 and 1, and $\boldsymbol{x^*}=(x^*_1,x^*_2,\dots,x_k^*)$ is the fixed point concentration of the cell model.
    We randomly switched the fitness function $f(\boldsymbol{x}^*;\boldsymbol{w^{(i)}})$ for $i=1,2,\dots,5$ every 10 generations and numerically evolved the reaction network of the cell.
    We adopted the environmental condition $\boldsymbol{s^{uni}}$ as $s^{uni}_i=1/n\ (i=1,2,\dots,n)$ with $n=10$.
    As shown in Fig.\ref{FIG:Non-growth}(a), maximum fitness within the population of the cells monotonically increased through evolution, although the growth rate of the fittest cells did not show the monotonic trend
    (Fig.\ref{FIG:Non-growth}(b)).
    In other words, fitness was not necessarily correlated with growth rate.
    Just as in the growth rate-based evolutionary simulations, only one singular value was separated from the others (see  Fig.\ref{FIG:Non-growth}(c)).
    This indicates that the choice of growth rate as an indicator of fitness was not the main reason for the one-dimensional constraint on phenotypic changes.
    
\newpage
\bibliography{hoge}
\bibliographystyle{unsrt}
\end{document}